\documentstyle[twoside,fleqn,espcrc2]{article}
\input prepictex
\input pictex
\input postpictex

\newdimen\xposition
\newdimen\yposition
\newdimen\dyposition
\newdimen\crossbarlength

\def\puterrorbar at #1 #2 with fuzz #3 {%
  \xposition=\Xdistance{#1}
  \yposition=\Ydistance{#2}
  \dyposition=\Ydistance{#3}

\setdimensionmode
\put {$\bullet$} at {\xposition} {\yposition}

\dimen0 = \yposition
  \advance \dimen0 by -\dyposition
\dimen2 = \yposition
  \advance \dimen2  by \dyposition
\putrule from {\xposition} {\dimen0}
  to {\xposition} {\dimen2}

\dimen4 = \xposition
  \advance \dimen4 by -.5\crossbarlength
\dimen6 = \xposition
  \advance \dimen6 by  .5\crossbarlength
\putrule from {\dimen4} {\dimen0} to {\dimen6} {\dimen0}
\putrule from {\dimen4} {\dimen2} to {\dimen6} {\dimen2}
\setcoordinatemode}

\newdimen\xpositionc
\newdimen\ypositionc
\newdimen\dypositionc
\newdimen\crossbarlength

\def\putcirclebar at #1 #2 with fuzz #3 {%
  \xpositionc=\Xdistance{#1}
  \ypositionc=\Ydistance{#2}
  \dypositionc=\Ydistance{#3}

\setdimensionmode
\put {$\circ$} at {\xpositionc} {\ypositionc}

\dimen0 = \ypositionc
  \advance \dimen0 by -\dypositionc
\dimen2 = \ypositionc
  \advance \dimen2  by \dypositionc
\putrule from {\xpositionc} {\dimen0}
  to {\xpositionc} {\dimen2}

\dimen4 = \xpositionc
  \advance \dimen4 by -.5\crossbarlength
\dimen6 = \xpositionc
  \advance \dimen6 by  .5\crossbarlength
\putrule from {\dimen4} {\dimen0} to {\dimen6} {\dimen0}
\putrule from {\dimen4} {\dimen2} to {\dimen6} {\dimen2}
\setcoordinatemode}

\newcommand{\AmS}{{\protect\the\textfont2
  A\kern-.1667em\lower.5ex\hbox{M}\kern-.125emS}}

\title{Gauge-Invariant Noncompact Lattice Simulations}

\author{Kevin Cahill\address{Department of Physics and Astronomy,
        University of New Mexico, \\
        Albuquerque, New Mexico 87131-1156, U.~S.~A.}%
                \thanks{Research supported
                by the U.~S. Department of Energy
                under contract DE-FG04-84ER40166;
                e-mail: kevin@cahill.unm.edu.}}

\begin{document}

\begin{abstract}
Three techniques for performing
gauge-invariant, noncompact lattice simulations of
nonabelian gauge theories are discussed.
\par
In the first method,
the action is not itself gauge invariant,
but a kind of lattice gauge invariance
is restored by random compact gauge transformations
during the successive sweeps of the simulation.
This method has been applied to pure $SU(2)$ gauge theory
on a $12^4$ lattice, and
Wilson loops have been measured
at strong coupling, $\beta=0.5$.
These Wilson loops display a confinement signal
not seen in simulations performed earlier with
the same action but without the
random gauge transformations.
\par
In the second method,
the action is gauge symmetrized
by integrations over the group manifold.
\par
The third method
is based upon a new, noncompact form of the
action that is exactly invariant
under lattice gauge transformations.
The action is a natural discretization of
the classical Yang-Mills action.
\end{abstract}

\maketitle

\section{INTRODUCTION}
In Wilson's formulation~\cite{Wils74}
of lattice gauge theory,
the basic variables
are the elements of the gauge group,
not the fields of the theory.
Using this formulation Creutz~\cite{Creu80a}
in 1980 displayed evidence for
confinement in both abelian and nonabelian gauge theories.
\par
The Wilson action is exactly invariant
under lattice gauge transformations
and reduces to the classical action
when the group elements
are close to the identity.
Wilson's elegant formulation
is ideal at weak coupling.
\par
But at strong coupling,
the group elements are
often far from the identity,
and the Wilson action is
quite different from the
classical Yang-Mills action.
It is unknown
whether Wilson's formulation is accurate
at the moderately strong couplings
where it is really needed.
Among the symptoms of trouble
are the false vacua~\cite{Cahi88} of the Wilson action
which at strong coupling
affect the string tension~\cite{Mack82,Grad88}
and the confinement signals
exhibited by Creutz~\cite{Creu80a}
in his simulations of
abelian gauge theories.
\par
To examine these questions,
some physicists have introduced lattice actions
that are noncompact discretizations
of the continuum action
with the fields as the basic variables.
For $U(1)$ these noncompact formulations are
accurate for all coupling strengths~\cite{Cahi86};
for $SU(2)$ they agree well with perturbation theory
at very weak coupling~\cite{Cahi89}.
\par
Patrascioiu, Seiler, Stamatescu, Wolff, and Zwanziger performed
the first noncompact simulations of $SU(2)$
by discretizing the classical action
and fixing the gauge~\cite{Patr81}.
They saw a Coulomb force.
\par
Later simulations~\cite{Cahi93}
were carried out with
an action free of spurious zero modes,
for which it was not necessary to fix the gauge.
The Wilson loops of these simulations showed
no sign of quark confinement.
A likely explanation of this negative result
is that these noncompact actions
lack an exact lattice gauge invariance.
\par
The first gauge-invariant noncompact simulations
were carried out by Palumbo, Polikarpov, and Veselov~\cite{Palumbo92}
and were based on earlier work by Palumbo {\em et al.}~\cite{Palumboet}.
They saw a confinement signal~\cite{Palumbo92}.
Their action contains five terms,
constructed from two invariants,
and involves auxiliary fields and
an adjustable parameter.
\par
The present paper discusses three
ways of performing gauge-invariant
noncompact simulations with an action
that is a discretization
of the classical Yang-Mills action
without extra terms or fields.
\par
In the first technique,
one subjects the fields
to random compact gauge transformations
during the Monte-Carlo updates~\cite{Cahi93}.
The random gauge transformations restore
a semblance of gauge invariance~\cite{CahLat93}.
This method has been used to measure
Wilson loops at strong coupling, $\beta\equiv 4/g^2=0.5$,
on a $12^4$ lattice
in a noncompact simulation of $SU(2)$ gauge theory
without gauge fixing or fermions.
In this simulation the Wilson loops fall off
exponentially with the area of the loop
at a rate that
is over two orders of magnitude greater
than the negligible rate seen in an earlier simulation
with the same action but without
the random gauge transformations.
\par
The second technique is to symmetrize
the action itself by subjecting it to a full set
of gauge transformations.  Here one inserts
a gauge transformation at each vertex
and integrates over the gauge group.
This technique is simpler to perform
if the gauge transformations are noncompact.
\par
The third method is based upon a new
noncompact action that is exactly
invariant under lattice
gauge transformations.
The action is a natural discretization
of the classical Yang-Mills action.
Three ways of interpreting this
action are outlined.
\section{THE FIRST METHOD}
\subsection{The Action}
In both the earlier simulations without
gauge invariance
and the present simulation
with random gauge transformations,
the fields are constant on the links
of length $a$, the lattice spacing, but are interpolated
linearly throughout the plaquettes.
In the plaquette with vertices $n$, $n+e_\mu$,
$n+e_\nu$, and $n+e_\mu+e_\nu$,
the field is
\begin{eqnarray}
A_\mu^a(x) & = & ({x_\nu \over a} -n_\nu) A_\mu^a(n+e_\nu) \nonumber \\
& & \mbox{} + (n_\nu +1-{x_\nu \over a})A_\mu^a(n),
\end{eqnarray}
and the field strength is
\begin{eqnarray}
F_{\mu\nu}^a(x) & = & \partial_\nu A_\mu^a(x)-\partial_\mu A_\nu^a(x)
\nonumber \\
& & \mbox{} + g f_{bc}^a A_\mu^b(x) A_\nu^c(x).
\end{eqnarray}
The action $S$ is the sum over all plaquettes
of the integral over each plaquette of the
squared field strength,
\begin{equation}
S=\sum_{p_{\mu\nu}}
{a^2 \over 2} \int \! dx_\mu dx_\nu F_{\mu\nu}^c(x)^2.
\label{action}
\end{equation}
The mean value in the vacuum of
a euclidean-time-ordered operator $Q$
is approximated by a ratio of multiple integrals
over the $A_\mu^a(n)$'s
\begin{equation}
\langle {\cal T} Q(A) \rangle_0  \approx
{\int
e ^ {-S(A)}  Q(A)
\prod_{\mu,a,n}  dA_\mu^a(n)
\over \strut
 \int
e ^ {-S(A)}
\prod_{\mu,a,n}  dA_\mu^a(n)
}
\end{equation}
which one may compute numerically.
Macsyma was used to write most of
the Fortran code~\cite{Cahi90}
for the present simulation.
\subsection{Gauge Invariance}
To restore gauge invariance,
the fields are subjected to
random compact gauge transformations
during every sweep, except those devoted
exclusively to measurements.
At each vertex $n$ a random
number $r$ is generated uniformly on the interval $(0,1)$;
and if $r$ is less than a fixed probability,
set equal to 0.5 in this work,
then a random group element $U(n)$
is picked from the group $SU(2)$.
The fields on the four links coming
out of the vertex $n$ are then subjected to the
compact gauge transformation
\begin{equation}
e^{-igaA_\mu^{\prime b}(n)T_b} =
e^{-igaA_\mu^a(n)T_a}U(n)^\dagger
\end{equation}
and those on the links entering the vertex
to the transformation
\begin{equation}
e^{-igaA_\mu^{\prime b}(n-e_\mu)T_b} =
U(n) e^{-igaA_\mu^a(n-e_\mu)T_a}.
\end{equation}
At weak coupling one should use
random gauge transformations
that are suitably close
to the identity.
\subsection{Wilson Loops}
The quantity normally used to study confinement
in quarkless gauge theories is the Wilson loop $W(r,t)$
which is the mean value in the
vacuum of the trace of a path-and-time-ordered
exponential of a line integral of the connection
around an $r\times t$ rectangle
\begin{equation}
W(r,t) = (1/d) \, \langle {\rm tr \> }
{\cal PT} e^{-i g\oint \! A_\mu^a T_a dx_\mu}
\rangle_0
\end{equation}
where $d$ is the dimension of the generators $T_a$.
Although Wilson loops vanish
in the exact theory~\cite{Cahi79},
Creutz ratios $\chi (r,t)$
of Wilson loops defined~\cite{Creu80b}
as double differences of logarithms of Wilson loops
are finite. For large $t$, $\chi (r,t)$ approximates
($a^2$ times) the force between a quark
and an antiquark separated by the distance $r$.
\par
In this simulation
the data are not yet sufficient to allow one to
determine the Creutz ratios beyond the $3\times4$ loop.
The Wilson loops therefore have been
fitted to an expression involving Coulomb, perimeter,
scale, and area terms.
\subsection{Measurements and Results}
It will be useful to compare this simulation
with an earlier one~\cite{Cahi93}
in which the fields were not subjected
to random gauge transformations.
Both simulations were done on
a $12^4$ periodic lattice with a heat bath.
The earlier simulation consisted of
20 independent runs with cold starts.
The first run had 25,000 thermalizing
sweeps at inverse coupling $\beta = 2$
followed by 5000 at $\beta=0.5$;
the other nineteen runs began at $\beta = 0.5$
with 20,000 thermalizing sweeps.
There were 59,640 Parisi-assisted~\cite{Pari83}
measurements, 20 sweeps apart.
\par
The present simulation with random gauge transformations
is very noisy.
It consists of 21 runs
of which 20 were from cold starts
and one from a random start.
Each began with 20,000 thermalizing sweeps.
After the first few hundred thermalizing sweeps
in each run,
the average value of the action was stable.
\par
Wilson loops have been measured every five sweeps
for a total of 1,313,202 measurements.
The values of the Wilson loops so obtained
are listed in the table.
The errors have been estimated
by the jackknife method,
with all measurements in bins of 4000
considered to be independent.
\par
The Wilson loops of the gauge-invariant
simulation fall off much faster with increasing loop size
than do those of the earlier simulation.
Because the data do not accurately determine
all the Creutz ratios, I have fitted both
sets of loops, including the non-diagonal loops,
to the formula
\begin{equation}
W(r,t) \approx e^{ a + b(t/r + r/t) - 2c\,(r + t) - d\, rt}
\end{equation}
in which $a$ is a scale factor,
$b$ a Coulomb term, $c$ a perimeter term,
and $d$ an area term.
For the simulation without random gauge transformations,
I found $a\approx 0.26$, $b\approx 0.20$,
$c\approx 0.39$, and $d\approx 0.00$.
For the simulation with random gauge transformations,
I found $a\approx 0.57$, $b\approx 0.15$,
$c\approx 0.51$, and $d\approx 0.18$.
In the gauge-invariant simulation,
the coefficient of the area-law term is
over two orders of magnitude larger than in the earlier
simulation which lacked gauge invariance.
\par
$$
\vbox{
\hbox{\it Noncompact Wilson loops at $\beta = 0.5$}
\vskip 1pt
\vbox{\offinterlineskip
\hrule
\halign{&\vrule#&
  \strut\quad#\hfil\quad\cr
height2pt&\omit&&\omit&&\omit&\cr
&${r\over a}\times{t\over a}$\hfil&&Not invariant\hfil&&
Invariant\hfil&\cr
height2pt&\omit&&\omit&&\omit&\cr \noalign{\hrule}
height2pt&\omit&&\omit&&\omit&\cr
&$1\times1$&&0.402330(6)&&0.254558(5)&\cr
height2pt&\omit&&\omit&&\omit&\cr \noalign{\hrule}
height2pt&\omit&&\omit&&\omit&\cr
&$1\times2$&&0.208096(5)&&0.082528(6)&\cr
height2pt&\omit&&\omit&&\omit&\cr \noalign{\hrule}
height2pt&\omit&&\omit&&\omit&\cr
&$2\times2$&&0.085426(4)&&0.018709(4)&\cr
height2pt&\omit&&\omit&&\omit&\cr \noalign{\hrule}
height2pt&\omit&&\omit&&\omit&\cr
&$1\times3$&&0.111792(5)&&0.027715(5)&\cr
height2pt&\omit&&\omit&&\omit&\cr \noalign{\hrule}
height2pt&\omit&&\omit&&\omit&\cr
&$2\times3$&&0.040008(3)&&0.005224(3)&\cr
height2pt&\omit&&\omit&&\omit&\cr \noalign{\hrule}
height2pt&\omit&&\omit&&\omit&\cr
&$3\times3$&&0.018080(2)&&0.001431(5)&\cr
height2pt&\omit&&\omit&&\omit&\cr \noalign{\hrule}
height2pt&\omit&&\omit&&\omit&\cr
&$1\times4$&&0.060451(5)&&0.009352(3)&\cr
height2pt&\omit&&\omit&&\omit&\cr \noalign{\hrule}
height2pt&\omit&&\omit&&\omit&\cr
&$2\times4$&&0.019212(2)&&0.001502(5)&\cr
height2pt&\omit&&\omit&&\omit&\cr \noalign{\hrule}
height2pt&\omit&&\omit&&\omit&\cr
&$3\times4$&&0.008517(1)&&0.000407(3)&\cr
height2pt&\omit&&\omit&&\omit&\cr \noalign{\hrule}
height2pt&\omit&&\omit&&\omit&\cr
&$4\times4$&&0.003993(1)&&0.000121(5)&\cr
height2pt&\omit&&\omit&&\omit&\cr \noalign{\hrule}
height2pt&\omit&&\omit&&\omit&\cr
&$1\times5$&&0.032726(5)&&0.003152(3)&\cr
height2pt&\omit&&\omit&&\omit&\cr \noalign{\hrule}
height2pt&\omit&&\omit&&\omit&\cr
&$2\times5$&&0.009269(1)&&0.000434(5)&\cr
height2pt&\omit&&\omit&&\omit&\cr \noalign{\hrule}
height2pt&\omit&&\omit&&\omit&\cr
&$3\times5$&&0.004041(1)&&0.000118(3)&\cr
height2pt&\omit&&\omit&&\omit&\cr \noalign{\hrule}
height2pt&\omit&&\omit&&\omit&\cr
&$4\times5$&&0.001889(1)&&0.000039(5)&\cr
height2pt&\omit&&\omit&&\omit&\cr \noalign{\hrule}
height2pt&\omit&&\omit&&\omit&\cr
&$5\times5$&&0.000893(1)&&0.000013(3)&\cr
height2pt&\omit&&\omit&&\omit&\cr \noalign{\hrule}
height2pt&\omit&&\omit&&\omit&\cr
&$1\times6$&&0.017717(4)&&0.001064(5)&\cr
height2pt&\omit&&\omit&&\omit&\cr \noalign{\hrule}
height2pt&\omit&&\omit&&\omit&\cr
&$2\times6$&&0.004474(1)&&0.000123(1)&\cr
height2pt&\omit&&\omit&&\omit&\cr \noalign{\hrule}
height2pt&\omit&&\omit&&\omit&\cr
&$3\times6$&&0.001919(1)&&0.000033(1)&\cr
height2pt&\omit&&\omit&&\omit&\cr \noalign{\hrule}
height2pt&\omit&&\omit&&\omit&\cr
&$4\times6$&&0.000896(1)&&0.000009(1)&\cr
height2pt&\omit&&\omit&&\omit&\cr \noalign{\hrule}
height2pt&\omit&&\omit&&\omit&\cr
&$5\times6$&&0.000423(0)&&0.000003(1)&\cr
height2pt&\omit&&\omit&&\omit&\cr \noalign{\hrule}
height2pt&\omit&&\omit&&\omit&\cr
&$6\times6$&&0.000201(0)&&0.000003(1)&\cr
height2pt&\omit&&\omit&&\omit&\cr}
\hrule}}
$$
\par
\begin{figure} [htb]
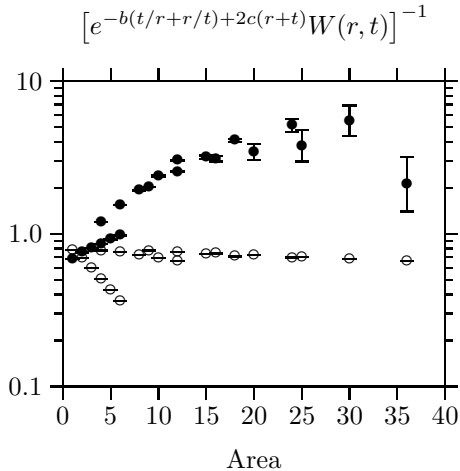

\beginpicture
\inboundscheckon
\setcoordinatesystem units <0.05in,0.8in>
\setplotarea x from 0 to 40, y from -1 to 1
\axis bottom label {Area} ticks
  numbered from 0 to 40 by 5 /
\axis left label {}
  ticks logged
  numbered at 0.1 1.0 10 /
  unlabeled short from 0.2 to 0.9 by 0.1
  from 2 to 9 by 1 /
\axis top label {$\left[ e^{-b(t/r+r/t)+2c(r+t)}W(r,t)\right]^{-1}$}
  ticks
  unlabeled from 0 to 40 by 5 /
\axis right ticks logged
unlabeled at 0.1 1.0 10 /
   short from 0.2 to 0.9 by 0.1
  from 2 to 9 by 1 /

\inboundscheckon

\crossbarlength=5pt

\puterrorbar at   1 -0.167179910 with fuzz  0.000008104
\puterrorbar at   2 -0.122458990 with fuzz  0.000029995
\puterrorbar at   4  0.077639264 with fuzz  0.000083329
\puterrorbar at   3 -0.093039574 with fuzz  0.000076462
\puterrorbar at   6  0.187240530 with fuzz  0.000264305
\puterrorbar at   9  0.305143866 with fuzz  0.001530045
\puterrorbar at   4 -0.065662357 with fuzz  0.000148588
\puterrorbar at   8  0.284095506 with fuzz  0.001408786
\puterrorbar at  12  0.406908302 with fuzz  0.003445306
\puterrorbar at  16  0.490568304 with fuzz  0.017784377
\puterrorbar at   5 -0.037832099 with fuzz  0.000448947
\puterrorbar at  10  0.378518214 with fuzz  0.004951851
\puterrorbar at  15  0.500507127 with fuzz  0.011707348
\puterrorbar at  20  0.534371070 with fuzz  0.051382816
\puterrorbar at  25  0.574531797 with fuzz  0.102330946
\puterrorbar at   6 -0.010726192 with fuzz  0.001926060
\puterrorbar at  12  0.480552568 with fuzz  0.003051459
\puterrorbar at  18  0.609911269 with fuzz  0.011714128
\puterrorbar at  24  0.710320751 with fuzz  0.042702826
\puterrorbar at  30  0.738330569 with fuzz  0.101200118
\puterrorbar at  36  0.324732948 with fuzz  0.178736716

\putcirclebar at   1 -0.108585959 with fuzz  0.000006153
\putcirclebar at   2 -0.159682852 with fuzz  0.000010560
\putcirclebar at   4 -0.110426483 with fuzz  0.000022826
\putcirclebar at   3 -0.227245654 with fuzz  0.000019929
\putcirclebar at   6 -0.118399289 with fuzz  0.000032347
\putcirclebar at   9 -0.110840366 with fuzz  0.000057889
\putcirclebar at   4 -0.297652054 with fuzz  0.000035489
\putcirclebar at   8 -0.137247834 with fuzz  0.000047242
\putcirclebar at  12 -0.121384319 with fuzz  0.000072399
\putcirclebar at  16 -0.129821869 with fuzz  0.000135925
\putcirclebar at   5 -0.368560540 with fuzz  0.000062234
\putcirclebar at  10 -0.158094889 with fuzz  0.000069811
\putcirclebar at  15 -0.134981643 with fuzz  0.000104236
\putcirclebar at  20 -0.142207037 with fuzz  0.000163179
\putcirclebar at  25 -0.154183587 with fuzz  0.000330575
\putcirclebar at   6 -0.439478042 with fuzz  0.000109557
\putcirclebar at  12 -0.179227384 with fuzz  0.000104815
\putcirclebar at  18 -0.148909414 with fuzz  0.000158418
\putcirclebar at  24 -0.155520382 with fuzz  0.000242348
\putcirclebar at  30 -0.166969061 with fuzz  0.000390073
\putcirclebar at  36 -0.180401319 with fuzz  0.000843690

\endpicture
\caption{The negative logarithms
of Wilson loops with the Coulomb and
perimeter factors canceled
are plotted against the area $rt$ of the loop.
The loops of the gauge-invariant simulation
are represented by bullets; those of the
earlier simulation by circles.}
\end{figure}
\par
To exhibit the renormalized quark-antiquark potential,
I have plotted in the figure the negative logarithms
$-\log_{10} \left[e^{-b(r/t+t/r)+2c(r+t)}W(r,t)\right]$ of the Wilson
loops with the Coulomb and perimeter terms removed.
Apart from the uncertain value of $W(6,6)$,
the loops of the gauge-invariant simulation,
represented by bullets, seem to display an area law;
whereas the larger loops of the earlier simulation,
represented by circles, show an essentially flat potential.
\section{THE SECOND METHOD}
While the random gauge transformations
of the first method
can restore a measure of gauge invariance,
it clearly would be better to use
an action that is itself exactly
gauge invariant or has been made so.
In the second method of performing
gauge-invariant noncompact simulations,
one symmetrizes an action that is not
itself gauge invariant.
To make the action gauge invariant,
one inserts a general gauge transformation
at each vertex of the lattice
and then independently integrates over
the group manifold at each vertex
using the Haar measure.
\par
One may take any suitable
noncompact action as a starting point,
for instance the action used
in the first method.
Since the action is positive definite,
the result of the integration over
the parameters of the group
will not be zero.
In practice one need only integrate over the
vertices of the plaquettes that contain
the arbitrary link that is to be updated,
fourteen in the case of the action (\ref{action}).
The relation between the gauge field
before and after the gauge transformation
is transcendental in the case of compact
gauge transformations and affine in the
case of noncompact gauge transformations.
Obviously the latter kind
would be easier to implement.
\section{THE THIRD METHOD}
The third method of performing
gauge-invariant noncompact simulations
is to start with a noncompact lattice
action that is intrinsically gauge invariant.
\subsection{Gauge Invariance}
Let us first decide what constitutes
a gauge transformation in this third method.
A good way to do that is to start
with the fermionic action density
which in the continuum is
\begin{equation}
\bar \psi ( i \gamma_\mu D_\mu - m ) \psi.
\end{equation}
A suitable discretization of the free part of that
action density is
\begin{equation}
{i \over a } \bar \psi (n) \gamma_\mu
[ \psi(n + e_\mu) - \psi(n)]
\end{equation}
in which $n$ is a four-vector
of integers representing an arbitrary vertex
of the lattice, $e_\mu$ is a unit vector
in the $\mu$th direction, and
$a$ is the lattice spacing.
The product of Fermi fields at the same
point is gauge invariant as it stands.
The other product of Fermi fields
becomes gauge invariant
if we insert a matrix $A_\mu(n)$ of gauge fields
\begin{equation}
{i \over a } \bar \psi (n) \gamma_\mu
[ (1 + i g a A_\mu(n)) \psi(n + e_\mu) - \psi(n)]
\end{equation}
that transforms under a gauge transformation
represented by the group elements
$U(n)$ and $U(n + e_\mu)$ in such a way that
\begin{equation}
1 + i a g A'_\mu(n) =
U(n) [ 1 + i a g A_\mu(n) ] U^{-1}(n + e_\mu).
\label{1+A'}
\end{equation}
The required behavior is
\begin{eqnarray}
\lefteqn{A'_\mu(n) = U(n) A_\mu(n) U^{-1}(n + e_\mu)}
\nonumber \\
& & \mbox{} + {i \over a g} U(n)
\left[ U^{-1}(n) - U^{-1}(n + e_\mu) \right].
\label{A'}
\end{eqnarray}
Under such a gauge transformation,
the usual gauge-field matrix
$A_\mu(n) = T_a A^a_\mu(n)$ remains
in the Lie algebra only to first order
in the lattice spacing $a$.
Three ways of coping with this problem
are outlined in Sec.~4.2.
\par
Let us define the lattice field
strength $F_{\mu\nu}(n)$ as
\begin{eqnarray}
F_{\mu\nu}(n) & = &
{ 1 \over a } [ A_\mu(n+e_\nu) - A_\mu(n) ]
\nonumber \\
& & \mbox{} - { 1 \over a } [ A_\nu(n+e_\mu) - A_\nu(n) ]
\nonumber \\
& & \mbox{} + i g  [ A_\nu(n) A_\mu(n+e_\nu)
\nonumber \\
& & \mbox{} - A_\mu(n) A_\nu(n+e_\mu) ]
\label{F}
\end{eqnarray}
which reduces to the continuum Yang-Mills
field strength in the limit $a \to 0$.
Under the aforementioned gauge transformation (\ref{A'}),
this field strength transforms as
\begin{equation}
F'_{\mu\nu}(n) = U(n) F_{\mu\nu}(n) U^{-1}(n + e_\mu + e_\nu).
\label{F'}
\end{equation}
The field strength $F_{\mu\nu}(n)$
is antisymmetric
in the indices $\mu$ and $\nu$, but it is not
hermitian.
To make a positive plaquette action density,
we write
\begin{equation}
{1 \over 4 k} {\rm Tr} [F^\dagger_{\mu\nu}(n) F_{\mu\nu}(n)],
\label{S}
\end{equation}
in which it is assumed that the generators $T_a$
of the gauge group have been orthonormalized as
${\rm Tr} ( T_a T_b ) = k \delta_{ab}$.
Because $F_{\mu\nu}(n)$ transforms
covariantly (\ref{F'}), this action density
is exactly invariant under the
noncompact gauge transformation (\ref{A'}).
\subsection{Three Interpretations}
In general the gauge transformation (\ref{A'})
maps the usual matrix of gauge fields
$A_\mu(n) = T_a A^a_\mu(n)$
into a matrix that lies outside the
Lie algebra of the gauge group.
I shall now outline three responses to
this problem.
\par
The first response is to note that
for group elements $U(n)$ of the form
\begin{equation}
U(n) = e^{-i a g \omega^a T_a },
\label{U}
\end{equation}
the gauge transformation (\ref{A'})
to lowest (zeroth) order in the
lattice spacing $a$,
does keep the gauge-field matrix in the
Lie algebra.  Thus one may use
the usual matrix $A_\mu(n) = T_a A^a_\mu(n)$
of gauge fields and retain
invariance under infinitesimal gauge
transformations.
\par
The second response is to accept
the fact that the gauge-field matrix
$A_\mu(n)$ will be mapped by the
gauge transformation (\ref{A'})
into a matrix that lies outside
the Lie algebra of the gauge group
and to use this more-general matrix
in the simulation.
Thus one may use the action (\ref{S})
in which the field strength (\ref{F})
is defined in terms of gauge-field
matrices that are of the
more general form
\begin{equation}
A_\mu(n) = V A^0_\mu(n) W^{-1}
+ {i \over a g} V \left( V^{-1} - W^{-1} \right)
\label{newA'}
\end{equation}
where $ A^0_\mu(n) $ is a matrix of gauge fields
defined in the usual way,
$ A^0_\mu(n) \equiv T_a A^{a,0}_\mu(n) $.
Here the group elements $V$ and $W$
associated with the gauge field $A_\mu(n)$
are unrelated to those
associated with the neighboring
gauge fields $ A_\mu(n+e_\nu) $,
$ A_\nu(n) $, and $ A_\nu(n+e_\mu) $.
I intend to test this method
in the near future.
\par
The third response is to
take a cue from the
transformation rule (\ref{1+A'})
and to represent the quantity
$ 1 + iga A_\mu(n) $ as a element $ L_\mu(n) $
of the gauge group.
In this case the matrix $ A_\mu(n) $
of gauge fields is
related to the link $ L_\mu(n) $ by
\begin{equation}
A_\mu(n) =  { \left[ L_\mu(n) - 1\right] \over iga }
\label{L}
\end{equation}
and the action (\ref{S})
defined in terms of the field strength
(\ref{F}) with gauge-field matrix (\ref{L}) is,
{\it mirabile dictu\/},
Wilson's action
\begin{eqnarray}
\lefteqn{S =}
\nonumber \\
& & \mbox{}
\!\!\!\!\!\!\!\!
{k - \Re \, {\rm Tr} L_\mu(n) L_\nu(n+e_\mu)
L^\dagger_\mu(n+e_\nu) L^\dagger_\nu(n)
\over 2 a^4 g^2k }.
\label{W}
\end{eqnarray}
\section*{ACKNOWLEDGMENTS}
I should like to thank
M.~Creutz for pointing out
a problem with an early version
of this work;
H.~Barnum, G.~Herling, G.~Kilcup, F.~Palumbo,
J.~Polonyi, D.~Topa, and K.~Webb
for useful conversations;
and the Department of Energy for support under grant
DE-FG04-84ER40166.
Most of the computations of this paper
were done on two DEC Alpha computers,
one of which was lent by the UNM neutrino group.
Computer time was also supplied by
by the UNM high-energy-physics group,
by IBM, by NERSC, and by Richard Matzner's
Cray Research grant at the
Center for High-Performance Computing
of the Univ.\ of Texas System.
Maple was used to check the identities (\ref{F'})
and (\ref{W}).


\begin{thebibliography}{9}
\bibitem{Wils74} K.~Wilson, {\it Phys.~Rev.~D\/} 10 (1974) 2445.
\bibitem{Creu80a} M.~Creutz, {\it Phys. Rev. D\/} 21 (1980) 2308;
                 {\it Phys.~Rev.~Letters} 45 (1980) 313.
\bibitem{Cahi88} K.~Cahill, M.~Hebert, and S.~Prasad,
                 {\it Phys.\ Lett.\ B\/} 210 (1988) 198;
                 K.~Cahill and S.~Prasad,
                 {\it Phys.~Rev.~D\/} 40 (1989) 1274.
\bibitem{Mack82} G.~Mack and E.~Pietarinen,
                 {\it Nucl.~Phys.~B\/} 205 (1982) 141.
\bibitem{Grad88} M.~Grady, {\it Z.~Phys.~C\/} 39 (1988) 125.
\bibitem{Cahi86} K.~Cahill and R.~Reeder,
                 {\it Phys.~Lett.~B\/} 168 (1986) 381;
                 {\it J.~Stat.~Phys.\/} 43 (1986) 1043.
\bibitem{Cahi89} K.~Cahill,
                 {\it Nucl.~Phys.~B (Proc.~Suppl.)\/} 9 (1989) 529;
                 {\it Phys. Lett.~B\/} 231 (1989) 294.
\bibitem{Patr81} A.~Patrascioiu, E.~Seiler, and I.~Stamatescu,
                 {\it Phys.~Lett.~B\/} 107 (1981) 364;
                 E.~Seiler, I.~Stamatescu, and D.~Zwanziger,
                 {\it Nucl.~Phys.~B\/} 239 (1984) 177 and 201.
\vfill\eject
\bibitem{Cahi93} K.~Cahill,
                 {\it Phys.~Lett.\/} B304 (1993) 307
                 and in {\sl The Fermilab Meeting, DPF '92}
                 edited by C. H. Albright,
                 P. H. Kasper, R. Raja, and J. Yoh
                 (World Scientific Publishing Co.\ Pte.\ Ltd., 1993),
                 Vol.\ 2, p.\ 1468.
\bibitem{Palumbo92} F.~Palumbo, M.~I. Polikarpov, and A.~I. Veselov,
                 {\sl Phys.\ Lett.\ \/} B297 (1992) 171\null.
\bibitem{Palumboet} F.~Palumbo, M.~I. Polikarpov, and A.~I. Veselov,
                 {\sl Phys.\ Lett.\ \/} B258 (1991) 189\null;
                 F.~Palumbo,
                 {\sl Phys.\ Lett.\ \/} B244 (1990) 55\null.
\bibitem{CahLat93} K.~Cahill,
                 {\sl Nucl.\ Phys.\ B (Proc.\ Suppl.)\/}
                 34 (1994) 231 and eprint hep-lat/9312077.
\bibitem{Cahi90} K.~Cahill,
                 {\it Comput.~Phys.\/} 4 (1990) 159.
\bibitem{Cahi79} K.~Cahill and D.~Stump,
                 {\it Phys.~Rev.~D\/} 20 (1979) 2096.
\bibitem{Creu80b} M.~Creutz,
                 {\it Phys. Rev. D\/} { 21} (1980) 2308;
                 {\it Phys.~Rev.~Letters} 45 (1980) 313.
\bibitem{Pari83} G.~Parisi, R.~Petronzio, and F.~Rapuano,
                 {\it Phys.\ Lett.\ B\/} 128 (1983) 418.
\end{thebibliography}
\end{document}